# Challenges in Adopting Companion Robots: An Exploratory Study of Robotic Companionship Conducted with Chinese Retirees

MENGYANG WANG, The Hong Kong University of Science and Technology (Guangzhou), China
KEYE YU, The Hong Kong University of Science and Technology (Guangzhou), China
YUKAI ZHANG, The Hong Kong University of Science and Technology (Guangzhou), China
MINGMING FAN*, The Hong Kong University of Science and Technology (Guangzhou), China and The Hong Kong University of Science and Technology, China

Companion robots hold immense potential in providing emotional support to older adults in the rapidly aging world. However, questions have been raised regarding whether having a robotic companion benefits healthy older adults, how they perceive the value of companion robots, and what their relationship with companion robots would be like. To understand healthy older adults' perceptions, attitudes, and relationships toward companion robots, we conducted multiple focus groups with eighteen retirees. Our findings underscore the social context encountered by older adults in China and reveal the mismatch between the current value proposition of companion robots and healthy older adults' needs. We further identify factors influencing the adoption of robotic companionship, which include individuals' self-disclosure tendencies, quality of companionship, differentiated value, and seamless collaboration with aging-in-community infrastructure and services.

CCS Concepts: • **Human-centered computing** → **Empirical studies in HCI**.

Additional Key Words and Phrases: Companion robot, Social assistive robot, Humanoid robots, Aging, Older adults

## 1 Introduction

Companionship is beneficial to older adults' well-being both cognitively and mentally [3, 21, 41]. Active cognitive activity and social contact are important factors of a healthy lifestyle[41]. However, it can be challenging for older adults to receive companionship as needed due to the increasingly imbalanced age demographic landscape. By 2050, the world's population of people aged 60 years and older will reach 2.1 billion, and the ratio of older adults to young people aged below 18 years old will reach 2.5:1 [65], which might lead to the shortage and high costs of professionals who

---

*Corresponding Author

Authors' Contact Information: Mengyang Wang, The Hong Kong University of Science and Technology (Guangzhou), Guangzhou, Guangdong, China, mengyangwang@hkust-gz.edu.cn; Keye Yu, The Hong Kong University of Science and Technology (Guangzhou), Guangzhou, Guangdong, China, keyeyu@hkust-gz.edu.cn; Yukai Zhang, The Hong Kong University of Science and Technology (Guangzhou), Guangzhou, Guangdong, China, yzhang118@connect.hkust-gz.edu.cn; Mingming Fan, The Hong Kong University of Science and Technology (Guangzhou), Guangzhou, Guangdong, China and The Hong Kong University of Science and Technology, Hong Kong SAR, China, mingmingfan@ust.hk.

can offer companionship [37, 76, 77]. While older adults' children and family members may provide companionship, it is not always available due to different life, work, or study schedules. Companion robots have been investigated as an alternative and show potential in improving the health and overall quality of life of older adults [4, 61, 68].

Early research on companion robots applies them to animal-assisted therapy as a substitution for real animals to reduce the loneliness of older adults. Bank et al.[4] demonstrated the effectiveness of using robotic dogs in animal-assisted therapy to reduce loneliness. Subsequent studies also find other positive effects of companion robots, including reducing stress, improving brain functioning, and forging social relationships [15]. Prior works on companion robots predominately focused on older adults living in long-term care facilities [45, 68] or with dementia [61, 69]. With the ever-severe aging problem and the COVID-19 pandemic, recent studies also investigated the use of companion robots for healthy older adults living in their own homes[36, 39]. It is important to note that in this context, the term "healthy older adults" refers to cognitively intact, independent-living older adults. For the purpose of conciseness and for all subsequent references, we will use this term to refer to such individuals.

Healthy older adults' perceptions and attitudes toward companion robots are underexplored. While some studies with healthy older adults show positive effects of companion robots in reducing loneliness with physiological measures [4, 68], other studies revealed negative feedback from older adults, such as the robot's capability not meeting their needs [29, 49, 57], "toy-like" appearance [23, 35, 60], potential infantilization and deception [7]. In addition, some studies suggested older adults prefer robots with service functions over companionship [18, 25, 47]. A recent study argued that companionship is not attractive enough to be a robot's primary function [96], a view which seems to be supported by the commercial failures of companion robots, such as Aibo [35]. Therefore it is unclear whether healthy older adults believe robotic companionship is beneficial. Meanwhile, prior work also shows mixed results regarding the perceived role of companion robots. While many studies identified the potential roles as assistants, friends, family members, pets, etc. [22, 53, 79], other studies show ambivalence in older adults perceived roles of companion robots [29]. Therefore, how healthy older adults perceive their relationship with companion robots is unknown, especially considering expected roles and ethical concerns, such as stigma, illusion, deception, etc.

To summarize, though many healthy older adults show a positive attitude toward having a robot companion, there is still insufficient understanding of why older adults perceive current companion robot designs as not meeting their needs. To determine whether and how companion robots should be designed for healthy older adults, and what features they should have, it is important to have a full understanding of their attitudes and perceptions towards robotic companionship, as well as their expected role for companion robots.

Motivated by these ongoing open-ended questions regarding companion robots for healthy older adults who lived in their own homes, we raised the following research questions (RQs):

- RQ1: What are their perceptions and attitudes towards robotic companionship?
- RQ2: What are the factors influence the adoption of companion robots?
- RQ3: How do they perceive the relationship between humans and companion robots in terms of role positioning and ethical considerations?

To answer the three RQs, we conducted six focus group discussions with 18 retirees, all aged above 50, who may possess the resources and inclination to be early adopters of companion robots. We recruited retirees as our participants, which includes individuals both over and under 60 years of age, representing the traditionally defined older adult population, as well as individuals who represent the future older adult population. This mixed background can provide a richer perspective for exploratory research. For conciseness in expression, we will refer to them as older adults.

The study began with participants completing a questionnaire on demographic information and design preferences for three appearance types, materials, and configurations. After that, participants interacted with the desktop robot Kebbi [1] to experience the robot's verbal and nonverbal capabilities powered by a large language model. Then, during a discussion session, participants discussed their perceptions and attitudes toward robotic companionship, rationales for supporting or rejecting companion robots, and their expected role of a companion robot in their lives.

Our findings contributed to the CSCW community by presenting the social context in which Chinese older adults encounter companion robots and highlighting the positive significance of companion robots in helping older adults face the transition from the traditional "three generations under one roof" family support model. Our study shown that healthy older adults hold an open and positive attitude toward companion robots as assistance in supporting aging-in-place while exhibiting different levels of need for additional companionship from robots. We identified five advantageous scenarios of robotic companionship and the underlying rationale to avoid the complex aspect of human relationships. We further explored the factors that influence the adopting of companion robots, which including individuals' self-disclosure tendencies, quality of companionship, differentiated value, the seamless collaboration with aging-in-community infrastructure and services. Those who has the tendency to adopt companion robots aimto get timely emotional responses and interactions to improve mood and quality of life rather than reduce loneliness. Additionally, we revealed the expected role of companion robots is rather an extract of primary features from human relationships and should not be interpreted entirely according to the same meaning in human society, and these roles could co-exist and shift as needed, the perceived role of robots was influenced by the quality of companionship as well. Furthermore, we uncovered that older adults were aware of the artificial nature of robotic companionship and regarded it as a choice of their own rather than a deception due to their relatively healthy status.

Based on these findings, we demonstrated companionship could be the primary function of robots. We discussed the necessity to pay attention to cognitively intact, independent-living older adults' emotional support needs. We proposed that a shift in value proposition was necessary to effectively promote companion robots to healthy older adults. We proposed design implications for high-quality robotic companionship. In summary, our study contributes to a deeper understanding of the perceptions and attitudes of healthy older adults toward companion robots.

## 2 Background and Related Work

### 2.1 Background

Existing works used different definitions for companion robots, therefore it is necessary to clarify the taxonomy debates and the definition we adopt. Some studies define companion robots as robots that primarily provide social interaction by acting as companions, without assisting with performing tasks [69]. Conversely, in other studies, companion robots are defined as robots that can conduct a variety of assistive functions [17, 19, 66, 95]. Looking back from a taxonomy perspective, companion robots belong to one of the two subgroups of socially assistive robots (SARs) together with service robots[12, 67], although few early studies viewed companion robots as a type of service robot [22, 33]. Companion robots focus on supporting psychological well-being [1], while service robots focus on supporting daily tasks. In this paper, we followed Kachouie et al [42] which argued that it is neither possible nor helpful to draw a solid line for such taxonomy, and we define companion robots as robots that provide companionship to users. We focus our discussion with participants on robotic companionship and do not exclude participants' comments on relevant assistive functions.

---

[1]https://moviarobotics.com/kebbi/

Research on companion robots for older adults began with the application of animal-assisted therapy as a substitution for real animals to reduce loneliness [52, 68]. To provide companionship, these robots are designed to be sociable and evoke a sense of uniqueness, belonging, and intimacy in the human-robot relationship. For example, the seal-like companion robot, Paro, is equipped with artificial fur and tactile sensors on its surface to simulate the tactile sensations experienced with pets [74]. Similarly, the robotic dog Aibo is capable of engaging in handshakes with people, akin to a pet dog, and its personality evolves through interaction with users [43]. Additionally, the cat-like robot iCat can convey basic facial expressions and emotions using its eyebrows, eyelids, mouth, and head position [83].

Previous research has extensively examined the sociability of SARs [10, 40], thus, it is essential to differentiate the terms sociability and companionship to clarify the scope of our research. Sociability and companionship are interrelated but possess distinct characteristics. SARs are expected to enhance "sociability between subjects or between subjects and other people" [1], and the effectiveness of sociability is commonly measured by social interaction, social behavior, engagement, communication and interaction stimulation, etc. In contrast, companionship aims to alleviate feelings of loneliness and being socially isolated, which is typically assessed by the loneliness and attachment scores. This paper will concentrate on older adults' home-based care settings, and focus on the needs and provision of companionship between humans and robots, rather than enhancing robot-mediated human-human social interaction.

## 2.2 Current barriers for companion robot

*2.2.1 Lack of need.* Many studies mentioned that healthy older adults perceive companion robots as suitable for people who are "older" or "lonelier" [50, 51, 57], one assumption is the lack of need for robotic companionship from healthy older adults' perspective. For instance, in a study conducted with healthy adults, only a small number of participants expressed a desire for a robotic pet as a result of loneliness while most participants reported not feeling lonely or isolated [49]. Other studies further raised questions on whether consumers would purchase social robots primarily for companionship or not[13, 35]. Previous studies have suggested that service functions were more valuable from older adults' views [18]. Recently some research demonstrated that companionship was not attractive enough to become the primary function of robots [24, 96]. For instance, Zuckerman et al. [96] conducted research with healthy older adults and revealed that the robot's primary function significantly affects the willingness of older adults to have it in their home, with companionship being perceived as an inappropriate primary function and therefore rejected by healthy older adults. In summary, the above studies indicate that healthy older adults do not perceive themselves as lonely or isolated, they prefer practical functions over companionship in robots and practical services, and they may not purchase robots with companionship as the primary feature. In contrast to these views, some other studies propose that healthy older adults hold a positive and open attitude toward companion robots. Therefore, it is crucial and yet to be explored whether healthy older adults actually have a demand for robotic companionship.

*2.2.2 "Toy-like" feelings.* Prior studies showed that healthy older adults perceive companion robots as "toys" that capture their interest at first but fail to sustain it as the novelty wears off quickly [23, 35, 60]. In the study of Moradi et al. [59], older adults described companion robots as devices that are enjoyable and charming but not practical. They further elaborated that they think of it as a toy if the robot fails to provide practical value. The research of Dereshev et al. [23] shows similar results, after participants(all aged below 50) lived or worked with Pepper, a humanoid social robot, for over six months, they noted that the robot's unchanging voice tone and random, incoherent actions, inhibiting any sense of attachment to them. Additionally, other studies further pointed out that the robot's limited functionalities also made it less interesting for participants. To sum up, previous studies mentioned several potential rationales why older adults might perceive companion robots as "toys" and difficult to build long-term relationships, such as lack of richness and unpredictability for conversation, and differentiation from both interaction experience and functionality perspectives. Hence, there is a need for further investigation into healthy older adults' perspectives regarding barriers to adopting companion robots.

2.2.3 *Stigma.* In addition to the assumption that there is no need for robotic companionship, another assumption pertains to the stigma held by older adults, who view companion robots as only suitable for individuals who are "older" or "lonelier" [50, 51, 57]. The literature suggests that when older adults view a device or product as for people who are older or frailer than they are, they perceive the stigma associated with these devices and stereotypical views of aging as decline and isolation [30, 63]. Lee et al. [50] revealed that most participants preferred robots designed to be used by anyone over assistive robots or entertainment robots, as they perceive entertainment robots as mere toys and assistive robots as associated with disability and aging stigma. Lee et al. proposed that robots should be designed to support older adults' successful aging, such as older adults' autonomy resilience, rather than assisting with potential disabilities. This viewpoint may be related to various factors such as cultural background, values, and cognitive status of older adults. Therefore, we took a step further to explore the assumption with healthy older adults in China as well.

2.2.4 *Concerns for human-robot relationship.*

*Perceived roles.* The expected role of robots is an important factor in developing their appearance and interactions [26], however, there is a scarcity of research that investigates the role of companion robots from the perspective of healthy older adults. When taking a broader perspective that includes both SARs and home robot categories, and not limiting the analysis to solely healthy older adults, previous research has commonly identified servant, employee, or tools as the most frequently mentioned roles. However, there is a divergence in the literature regarding whether users perceive these robots as having more intimate roles or not, such as friend or peer or buddy, and pets[22, 53, 79]. For example, Ling et al. investigated the potential roles of SARs with a sample of 24 individuals aged 50 and above and found that the most frequently mentioned roles were servant or employee, and pet, followed by friend/ peer, and tool. Some participants expressed a desire for robots to fulfill multiple roles [53]. Similarly, Tobis et al. explored the expected roles of humanoid service robots from the perspectives of healthy older adults and their caregivers and discovered that, in comparison with their caregivers, older adults gave higher assessment scores to the view that the robot should serve as a companion and assistant for older adults [79]. Additionally, Chen et al. conducted a study with older adults suffering from depression and mild cognitive impairment who resided in long-term care facilities, and over a long-term trial with Paro, identified that older adults perceived Paro as a friend [15].

On the contrary, several other studies have indicated that individuals tend to perceive robots as tools rather than potential companions. For instance, Frennert et al. conducted a mixed-method study involving healthy older adults and found that participants expressed skepticism during interviews about forming meaningful relationships with robots [29]. The potential reason for this skepticism was identified as the stigma associated with relying on a machine for companionship, which may signal to others that they are lonely and fragile. In contrast, having a robot as a servant was perceived as acceptable and satisfactory. Similarly, Dautenhahn et al. investigated the role of companion robots with 28 adults under the age of 55 through questionnaires and human-robot interaction trials and found that most participants viewed companion robots as assistants, machines, or servants, with few expressing a desire for a robot companion to be a friend [22].

Hence, there is a lack of clarity regarding how healthy older adults perceive the role positioning of companion robots. Furthermore, there is uncertainty regarding the similarities and differences between the roles ascribed to robots, such as friends, family members, and pets, and their counterparts in real-world human relationships.

*Ethical concerns.* Among the six primary ethical concerns surrounding the use of companion robots for older adults identified by Körtner et al.[48], this study will specifically focus on the ethical issues of deception and isolation from the perspective of healthy older adults, as the remaining four issues are either particularly pertinent to specific subgroups (dignity and vulnerability issues for older adults with dementia or disabilities) or applicable to general robot category (privacy and safety issues).

There is debate regarding whether robotic companionship constitutes deception. Many researchers contend that the sociability of robots, which fosters an inappropriate illusion, constitutes a form of deception, as these robots tend to feign qualities that they do not possess [48, 84, 88]. Wallach's work, for instance, highlights how the sociability design of robots, which enables them to detect and respond to social cues, can be considered a type of deception [86]. Furthermore, Sharkey et al. have argued that deception can occur in social robotics, whether intentionally or not [73]. Turkle's study demonstrated a significant emotional attachment was observed between older adults and companion robots, noting that older adults felt frustration if the robots were not programmed to acknowledge their attention by saying their name [82]. Moreover, the recent advancement of Large Language Models (LLM) technology has significantly improved the intelligence and naturalness of speech interaction [14]. This progress further blurs the distinction between artificial intelligence and human interlocutors from a dialogue perspective. Conversely, other studies have emphasized it does not constitute deception as there is an absence of intention to deceive people into believing a robot is a genuine human or animal, furthermore, it is rarely the case that people are fooled into believing so [20, 75]. Yamazaki et al conducted a long-term study on companion robots involving 13 older adults living in their own homes. The study addressed ethical criticisms of deception by robots and found that both the older adults and their relatives prioritized the importance of the relationship with the robots and the benefits they received from them over the artificial attribute of robotic companionship [91]. These findings align with those of Bradwell et al [8], suggesting that initial ethical concerns may be justified if interactions with robots are beneficial in stimulating or soothing relatives, or in easing challenges faced by family members.

Aside from the concerns of deception, the research of Bates et al further questioned whether companion robots were justification to leave older adults alone for longer periods [5]. However, Lucidi et al attributed social isolation as a result of social and generational factors rather than as a consequence of using companion robots [6]. Meanwhile, many of the aforementioned viewpoints are not directly derived from empirical studies of older adults but rather reflect researchers' views of potential issues. Therefore, the direct perspectives of healthy older adults on this topic require further exploration and supplementation. Investigation of healthy older adults' perceptions is needed to gain a comprehensive understanding of this complex topic.

In summary, very few prior works on companion robots in the aging context focused on healthy older adults. From the previous research, companion robots had been questioned by healthy older adults for mismatching with their needs, the difficulty of main long-term engagement, and the controversial human-robot relationship. Inspired by those barriers from previous research, our research aims to concentrate on healthy older adults' perception and attitude toward companion robots and answer the question of whether those barriers come from older adults' resistance to the nature of robotic companionship or the design and presentation of robotic companionship, and whether there is still any design space for companion robots aimed to serve healthy older adults.

## 3 METHOD

We conducted six participatory focus groups, each with three retirees, to understand their perceptions and attitudes toward companion robots. The study received approval from the university ethics review board.

### 3.1 Participant

We recruited 18 retirees, aged from 50 to 72 years old (M=60.2, SD=7.4), with four males and fourteen females. We recruited participants aged 50 and above, a criterion that is aligned with the research of Tang et al.[78] and mirrors the legal retirement ages in China, which are set at 60 for men, 55 for female civil servants, and 50 for female workers[89]. This criterion was chosen not only to align with the official retirement ages but also to acknowledge that retirement represents a crucial life transition that profoundly influences individuals' lifestyles and well-being. Most of them gained degrees from senior high school (N=6) or technical college (N=4). All older adults lived with their spouses, and none of them had previous experience with companion robots. We advertised our recruitment information at a local continuing education school for older adults. The participants in this study exhibited a moderate to high socioeconomic status, suggesting that they may possess the resources and inclination to be early adopters of companion robots. The focus group discussions were audio-recorded, and translated for further analysis. Table 1 shows participants' demographic information. Besides age, gender, and education background, we also designed a multiple-choice question to gain information about participants' access to technology, specifically exploring their daily use of social media apps (like TikTok), online shopping apps (like Taobao), smart home appliances (like smart speakers, sweeping robots), and so on.

### 3.2 Stimuli

To provide participants with a better understanding of the potential forms of companion robots and to give them a tangible experience of the sociable design of robots, we prepared pictures of representative robots along with a desktop robot named Kebbi. Participants were also reminded not to limit themselves to the design forms and capabilities presented in the stimuli.

In the picture stimulus, we classified potential forms of companion robots into three types. For robots that are neither animal-like nor humanoid, some studies name them "object-like robots" or "non-biological form robots" [17], which refer to robots with abstract shapes that do not resemble animals or humans, albeit exhibiting certain social characteristics, such as facial expressions. Given the term "object-like" is relatively abstract and might bring confusion to older adults, the terminology "non-biological form robots" was adopted in this study.

In the demonstration phase, we utilized a desktop robot named Kebbi, which features a 7-inch multi-touch display panel, 12 free-moving joints, and five sensor regions, enabling autonomous movement. The original Kebbi is proficient in natural language interaction in Mandarin and Cantonese, as well as facial expressions and body gestures using hands, arms, and the head. In addition, we enhanced the voice interaction by integrating the advanced language model ChatGPT 3.5 to enhance the naturalness and coherence of voice interaction. We presented Kebbi to provide participants with a tangible experience of robots' nonverbal cues, as well as to showcase the current voice interaction capabilities empowered by the large language model. Figure 1 summarizes the key features we showed to participants.

### 3.3 Procedure

The study took place in meeting rooms at a university during July to August 2023, and lasted approximately 60 minutes for each group, Figure 2 summarizes the whole procedure. The focus group discussions were audio-recorded, and translated for further analysis. Participants first signed the consent form and then completed a questionnaire that gathered demographic information and design preferences for three appearance types, materials, and configurations. Afterward, the moderator showed the first stimulus, a paper that displayed representative companion robots of various forms. Next, the moderator presented Kebbi and demonstrated its capabilities of voice interaction, facial expression, body gestures, autonomous movement, etc. Participants were encouraged to interact with Kebbi after the moderator's introduction.

Table 1. Participants' demographic information.

| ID | Group no. | Age | Gender | Education background | Access to technology |
|----|-----------|-----|--------|----------------------|----------------------|
| P1 | Group 1 | 54 | Female | Undergraduate college | Social media app, information app, online shopping app, smart appliances |
| P2 | Group 1 | 56 | Female | Undergraduate college | Social media app, information app, online shopping app, smart appliances |
| P3 | Group 1 | 52 | Female | Technique college | Social media app, information app, online shopping app |
| P4 | Group 2 | 55 | Female | Senior high school | Social media app, information app, online shopping app, smart appliances |
| P5 | Group 2 | 50 | Female | Technical secondary school | Social media app, information app, online shopping app |
| P6 | Group 2 | 56 | Female | Technical secondary school | Social media app, information app, online shopping app |
| P7 | Group 3 | 66 | Female | Technical secondary school | Social media app, information app, online shopping app |
| P8 | Group 3 | 68 | Male | Senior high school | Social media app, information app, online shopping app |
| P9 | Group 3 | 61 | Female | Senior high school | Social media app, information app |
| P10 | Group 4 | 51 | Female | Junior high school | Social media app, information app, online shopping app |
| P11 | Group 4 | 57 | Male | Senior high school | Social media app, information app |
| P12 | Group 4 | 53 | Female | Junior high school | Social media app, information app, online shopping app |
| P13 | Group 5 | 70 | Male | Technique college | Social media app, information app, smart appliances |
| P14 | Group 5 | 69 | Female | Technique college | Social media app, information app, online shopping app |
| P15 | Group 5 | 60 | Female | Technique college | Social media app, information app, online shopping app |
| P16 | Group 6 | 72 | Female | Undergraduate college | Social media app |
| P17 | Group 6 | 66 | Female | Senior high school | Social media app, information app, online shopping app |
| P18 | Group 6 | 68 | Male | Senior high school | Social media app, information app |

At the beginning of the discussion session, participants were reminded that they should not limit their later discussion to specific forms or capabilities that were shown before. During the discussion session, participants were asked about their general understanding of companionship, their perception of robotic companionship, and their needs. The study also investigated participants' attitudes towards companion robots, including their rationales for supporting or rejecting them and whether they viewed it as a form of deception. Finally, participants were asked to consider the ideal role that a companion robot should play in their lives and provide their rationales.

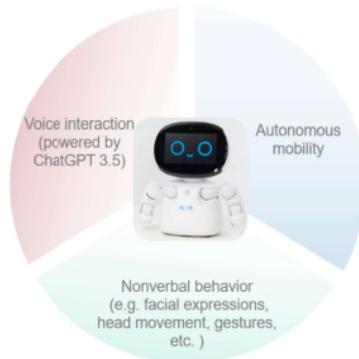

Fig. 1. Functionalities demonstrated by Kebbi, 1) Voice interaction powered by ChatGPT3.5; 2) Nonverbal behavior, such as facial expressions, head movement, and gestures; 3) Autonomous mobility.

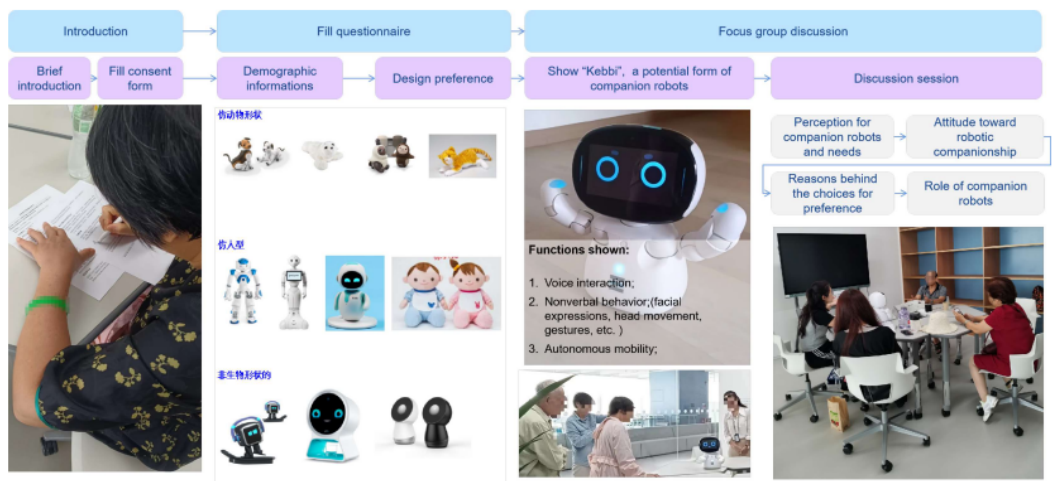

Fig. 2. The procedure of the focus group discussion: 1) Introduction phase: a brief introduction by the moderator, fill consent form by participants; 2) Questionnaire phase: fill demographic information and design preference by participants assisted with picture stimulus that displayed representative companion robots of various types, including humanoid robots, animal-like robots, and non-creature-style robots; 3) Focus group discussion phase: firstly show voice interaction, nonverbal behavior(facial expressions, head movement, gestures, etc.) and autonomous mobility of robot Kebbi as an example of companion robots by the moderator, then discussion phase started.

### 3.4 Data analysis

The research adopt thematic analysis[9] as a method for data analysis. All focus group discussions were conducted in Mandarin. The focus group discussion were recorded and subsequently transcribed verbatim. Then, two of the researchers independently coded the script using an open-coding approach[44]. They met regularly to collaboratively review and discuss their codes. A third researcher joined them to challenge the codes and resolve instances of disparity in coding through a comprehensive articulation of individual coding rationales, followed by a deliberative process aimed at conflict resolution. Through these means, a consensus was ultimately achieved, leading to the consolidation of the code inventory. Subsequently, the three researchers undertook affinity diagramming as a method to cluster the codes, thereby identifying overarching themes that emerged from these grouped codes. Moreover, an effort was made to amalgamate overlapping themes into cohesive clusters, enhancing the thematic coherence of the presentation. We present the themes and corresponding codes, along with supporting quotes,in the next section.

## 4 Findings

We present our key findings on three themes, 1) Social context for companionship; 2) Robotic vs human companionship; 3) Utility of companion robots. The first theme "Social context for companionship" underscores the social context encountered by Chinese retirees, including family structure change, and the desire for high-quality aging-in-place, and ethical concerns. The second theme "Robotic vs. human companionship" reveals retirees' perception of the difference between robotic companionship and human companionship, the human-robot relationship from the expected role perspective, and ethical concerns. The third theme, "Utility of companion robots" underscored the practical value of companion robots by exploring how retirees perceive whether companionship can serve as the primary function. Additionally, it examined the dual approaches of companion robots to deliver companionship and further revealed the key factors that influence the adoption of robotic companionship.

Section 4.1 is intended to unveil the background of RQ1, while the sub-themes "Companionship could serve as a primary function" and "Dual approaches of delivering companionship" in sections 4.3.1 and 4.3.2, respectively, aim to address RQ1 by exploring participants' perceptions. The second theme in section 4.2 is designed to answer RQ3, focusing on the perception of human-robot relationships. Furthermore, the sub-theme "Factors influencing the adoption of robotic companionship" in section 4.3.3 aims to respond to RQ2.

### 4.1 Social context for companion robots

*4.1.1 Technology-assistance to prolong independent living period.* Almost all participants expressed a desire for aging in place and openness to technology assistance. We found participants widely acknowledged technology as an effective aid for independent living and believed that technology-assisted aging is the prevailing trend, driven by factors such as the rising cost of human care resulting from low birth rates, the changing traditional family structure caused by rapid urbanization and the one-child policy, and their confidence in technological advancements. Firstly, participants assumed that in the future, they would have to rely more on technological assistance rather than manual assistance, as the declining birth rate would reduce the pool of young caregivers and consequently increase the costs associated with manual caregiving. As P16 explained, *"The population is aging so quickly, and young people are not having children. In the future, we will need robots to address the aging population issue."* Secondly, participants mentioned that family nuclearization has made it difficult to rely on their own children for eldercare, as opposed to the way the previous generations did. As they are the first generation impacted by the one-child policy, such concerns become particularly acute. Meanwhile, the rapidly developing society has placed immense competitive pressure on the younger generation, limiting their energy to provide physical and emotional care. Apart from the predicted shift towards technology-assisted aging, driven by the potential scarcity of both paid manual assistance and familial support from their children, participants maintain a profound faith in technological advancements, having witnessed China's remarkable social progress and technological advancements. All these rationales lead to an open and positive attitude toward technology-assistance to prolong their independent living period.

*4.1.2 Technology-assistance to reduce children's burden.* Furthermore, we revealed that participants held the mindset that maintaining independent living capabilities in the Chinese social context also means not burdening their children and relatives with additional concerns, beyond merely considering personal autonomy and dignity. Participants elaborated on the tension they felt between their eagerness to connect with their children, the busy schedules of their children, and their conscious effort to restrain excessive contact to avoid causing undue disturbance. As P1 underscored, *"When our children are in their thirties and forties, it is a time of high pressure and busyness. We have all been through that stage, so at that time, taking care of ourselves and not causing trouble for them is the greatest help we can provide."* After a lifetime of supporting their children, Chinese parents tend to believe that once their offspring are settled in their careers and they themselves are aging, avoiding causing trouble for their children is the most effective form of 'help' they can offer. Therefore, utilizing technological solutions is viewed as a strong pillar for maintaining independent living capabilities, and having robotic companionship is perceived as being able to provide peace of mind for their children, as P1 explained, *"Sometimes, our children also worry about how we are at home. With a robotic companion, they'll feel more reassured about our situation."*

### 4.2 Robotic vs human companionship

*4.2.1 Robotic companionship can be a complement to human companionship.* While acknowledging human or animal companionship cannot be replaced, participants pointed out scenarios in which they preferred robotic companionship as it could avoid human nature's complexity. First, Participants hold the view that human or animal companionship is irreplaceable, due to their authentic nature as well as the richness and unpredictability of the interactions they provide, and thus it is preferred for the majority of the time. As P11 explained, *"Just like a pet dog, if you hit it, it might bark, if you scold it, it might just roll on the ground, its response is not programmed."* However, participants also mentioned several scenarios where they found the subjectivity and complexity aspects of human nature to be barriers to companionship, and therefore they turned to prefer robotic companionship instead. We categorized those scenarios into five types.

**Too verbose to share with others.** Persistently sharing feelings and gossip with loved ones could potentially evoke irritation. P4 emphasized: *"I can speak more than thirty thousand sentences every day. My husband will feel bothered if I talk to him all the time...Companion robots offered me an outlet for uninhibited emotional expression."*

**Too tough or personal to disclose to others.** Robots could be perceived as more trustworthy confidants compared with humans, due to privacy concerns. Specifically, P9 elaborated: *"It could safeguard the sanctity of my personal information, but people may slip out it."*

**Too vulnerable to potential conflicts.** The unavoidable conflict of opinions or emotions in human conversations might contribute to inadequate emotional solace. Participants expressed that when they feel tired, they often desire comfort. However, due to insufficient understanding, differences in perspectives, or conflicting interests, human companions may get caught up in analyzing and judging the situation, which in turn makes them feel even more exhausted. P15 underscored: *"Companion robots could offer me unwavering emotional support."*

**Too vulnerable to potential risks.** Participants also mentioned the credibility and reliability of companion robots were appealing particularly in contexts involving vulnerable individuals like children or older adults. P9 gave an example: *"If I employ a human servant to take care of myself, the duplicity may exist. However, robots will always follow my commands."*

**Too narrow to explore broadly.** Participants explained another typical scenario to use companion robots, which utilized companion robots' distinctive capacities to furnish diverse perspectives, contemporary insights, and innovative solutions that stood in contrast to the limitations of human perspectives. In addressing specific issues, due to the similarity of living environments, mindsets, and knowledge backgrounds, suggestions from one's surroundings are often limited. On the other hand, companion robots can leverage their extensive knowledge base and diverse perspectives to provide participants with novel insights and solutions from different fields, broadening their horizons and increasing their possibilities in problem-solving. P6 specified: *"Robots could give me multiple ideas from different areas, and even help me acquire knowledge."*

To summarize, while human companionship is inevitably fraught with the subjectivity of individual personalities and the complexity of human nature, the companionship of robots offers a more objective and uncomplicated experience. Participants felt there was no need to debate with or persuade robots, nor did they have to guess or worry about being judged by robots or about robots acting against their own interests. Therefore, robots' perceived qualities, such as simplicity, trustworthiness, and reliability, make them an appealing choice under the aforementioned scenarios.

*4.2.2 Expected Roles.* Regarding the roles assigned to companion robots, participants generally categorized them into three domains: assistants or servants, friends, and family members. Notably, most participants ((N=12) chose assistant or servant, followed by identifying the role of friends (N=4), with only two participants considering them as family members. **1) Servant or assistant.** There is a notable overlap between participants who perceived companion robots as assistants or servants and those who earlier expressed the preference for service functions that could support independent living. P7 pointed out: *"I think companion robots should be servants who could do some household chores."*; **2) Friend.** Participants who categorized companion robots as friends predominantly emphasized the importance of "emotional companionship." Their perspective centered on the idea that interacting with these robots evoked a sense akin to companionship with friends. P10 further explained, *"Equitable exchange is important in friendship. Robots treat me sincerely just like my good friend."* However, the requirements to "follow my commands" have also been mentioned by those participants; **3) Family member.** Two distinct rationales explain why participants considered companion robots as family members. The first one is the absolute loyalty these robots exhibit, which encourages open communication similar to what one has with family. The second rationale is that, as participants received limited contact from their children and grandchildren, companion robots could alternatively take this role to fill the void of familiar presence. P2 precisely articulated, *"The companion robot in my home is like my young son."*.

Apart from these common roles, participants also pointed out that multiple roles could co-exist and the roles could switch according to the quality of companionship. Participants underscored companion robots can switch between multiple roles, including offering emotional support like friends and providing services like reminders similar to assistants or servants. P13 noted, *"The seamless switch between roles is similar to how humans often take on various roles, and this adaptability was not seen as problematic, at least for me, as long as it suited the specific context."* Consequently, there was no conflict in having these robots fulfill multiple roles simultaneously. Participants also mentioned the nature of companion robots served as tools, as P12 elaborated earlier. Nevertheless, they indicated that the potential for forming relationships with companion robots depends on the quality of the companionship experience those robots could provide rather than their tool-like attributes. P13 explained, *"If the companionship is good enough, the attachment will be build up gradually and naturally."*

When describing companion robots with these roles, participants employed specific prominent features of human social relationships to express their expectations, rather than equating the human-robot relationship with their counterparts in human society. For instance, while a group member positioned a companion robot as a family member, P13 pointed out that *"There is a strong connection between family members and shared genetic heritage, including the ability to pass down an inheritance, but it is crazy and unrealistic to pass my legacy to a robot."* Similarly, when a group member shared the idea to take a companion robot as a friend, P12 questioned, *"A friend is an equal relationship, you can't order it around. So, it is not a friend but an assistant; no matter how good the company is, it is still a tool."* The skepticism raised and subsequently acknowledged by the group members during the discussion represents the notion that, when describing these roles, participants only extracted a portion of the characteristics of these roles to depict what they perceive as the most prominent features of companion robots.

*4.2.3 Ethical concerns.* We present ethical concerns in two aspects, the perception of whether there is deception or not, and attitudes toward other potential negative influences arising from the use of companion robots.

Though robotic companionship was generated through software, participants emphasized that as they were in relatively good cognitive health status, it was a conscious choice rather than a deceptive experience for them. They chose it knowingly and made a deliberate decision based on understanding its functions and principles. The selection process is transparent, unrestricted, and self-directed. P7 mentioned, *"It does not make any sense to debate deception as there is no financial loss involved; instead, the outcomes are companionship and emotional contentment."* Similarly, P15 further elaborated, *"Companionship from robots is similar to services offered by psychotherapists. Robots aim to fulfill user needs and provide genuine solace, much like psychotherapists, they serve us sincerely without hidden deception."* Consequently, the outcome of the companionship process is intrinsically aligned with the objective of affording companionship to users, thus nullifying the notion of deception. To summarize, robotic companionship is merely perceived as a problem-solving tool deployed by their independent decision.

Furthermore, regarding potential negative influences caused by the usage of companion robots, participants articulated their willingness to assume accountability for the outcomes of their decisions as they were healthy and conscious. Nevertheless, participants also emphasized the need for regulations to ensure that companion robots are designed with positive values, particularly considering potential users with disabilities or severe depression. As P17 elaborated, *"For healthy people, expressing their feelings and going through this process is not a problem. However, those who are already emotionally fragile and on the verge of a breakdown, may misinterpret things and end up exacerbating their issues"*. Moreover, participants suggested that the design of the robot could help mitigate potential emotional projection, transference, and illusion. For example, robots should not resemble or imitate familiar intimate persons on purpose, or as P7 mentioned, *"I prefer a companion robot with a hard shell. I'm getting old, and I hope that touching a hard shell will remind me that this is not real."*

### 4.3 Utility of companion robots

*4.3.1 Companionship could serve as a primary function.* Though previous work argued companionship could not be a primary function for robots due to its low attractiveness compared with other service functions, participants in our research hold the view that robotic companionship is a functionality that in parallel to other assistive services. From participants' point of view, robotic companionship has its own values and can exist either as an independent primary feature of a robot or can also be integrated with other functionalities. Besides, retirees anticipate that as the industry of social robots for the aging population matures, there will be abundant social robot offerings in the market, which is similar to the diverse consumable electronic products nowadays. As such, P16 elaborated: *"If robots could handle cooking duties, there is potential for them to also offer companionship ...... Integration with other functions is also OK; however, it makes the product complex and expensive."* Therefore, it's a matter of product definition and portfolio design to balance perceived value and price, rather than resistance to companionship function due to its low significance.

*4.3.2 Dual approaches of delivering companionship.* Though the research field divides social assistive robots into service robots and companion robots, participants mentioned that the line between companionship and service functions can become blurred sometimes, both typical emotional companionship and service could be the approach to deliver companionship. Service functions typically pertain to utility-oriented tasks, such as assisting with daily living activities, while companionship emphasizes emotional support and the sense of being accompanied. First, we found participants considered companionship can be the purpose, while services could be the approach to achieving it. For example, P13 mentioned, *"Companionship is a big combination, which includes caring, presence with your hobbies, communication, and taking you out to have fun, all of these could fall under companionship."* Second, when designed with intention, service-oriented functions can impart not just utility but also a profound sense of being cared for and accompanied. As P2 mentioned, *"When I quarrel with my husband, the companion robot could remind me 'Hey, calm down, be careful of your blood pressure.'* Although blood pressure monitoring is a service-oriented function, designing it in the way P2 described delivers a strong sense of being cared for. Therefore, it could also be identified as a companion-oriented function. Furthermore, this ambiguity in categorization was particularly evident in the context of health and safety-related services. Services regarding health and safety could deliver older adults the feeling of being noticed and cared for, thus evoking the feeling of being accompanied. Participants envisioned various scenarios to demonstrate this view, for example, offering cognitive training and warnings for cognitive decline, environmental safety guards (such as timely alerts for toxic or hazardous risks like carbon monoxide in a room), and providing water or massage when they experienced fatigue or low mood. As explained by P15 *"even a small pat on the back during distress can be a form of comfort."* Those scenarios are highly relevant to health and safety, suggesting that these services are particularly effective in conveying companionship. In conclusion, the blurred distinction between companionship and service functions highlights that companionship can be expressed through emotional support, as well as through the provision of services.

*4.3.3 Factors influencing the adoption of robotic companionship.* Divergent perspectives emerged about the willingness to adopt robotic companionship in the current life stage, indicating the perceived utility of robotic companionship varies among different participants. The adoption willingness could be further categorized into three types: support (N=4), neutral (N=6), and reject (N=8). Aside from pragmatic limitations like affordability and spatial compatibility, we identify several key factors that play an important role in participants' choices.

*Individuals' self-disclosure tendencies.* First, we found individuals' self-disclosure tendencies difference might influence their need for extra companionship. Those who chose to adopt companion robots expressed they had a vibrant social life, not experiencing loneliness or isolation. They choose companion robots in order to have a partner who can constantly observe, understand, and provide emotional responses, thus satisfying their tendencies for self-disclosure and receiving responses, and enabling them to have another way of self-expression and interaction. As P2 elaborated, *"I have a quite busy schedule every day with painting, doing sports, and attending instrument courses with my friends"*. On the contrary, Some participants in the reject cohort demonstrated a clear inclination for no need for additional robotic companionship currently or in the future. For instance, P13 expressed his introverted and emotionally composed nature, and emphasized his ability to manage and even relish solitude, suggesting that those participants could cope with "staying alone" in various ways.

*Quality of companionship.* Companionship is a complex interplay that involves recognition, understanding, and uplifting emotions, resulting in a profound shared presence.During the focus group discussion, we found that attaining high-quality companionship is a challenging feat and existing robots might fail to provide expected companionship. We discovered the quality of robotic companionship is influenced by diverse capabilities such as the perception of users' emotional states, accurate judgments of current situations, diverse ways of expressing emotional support, and the ability to select appropriate expressions based on suitable contexts. First and foremost, participants emphasized the importance of tailored strategies, transcending the limitations of a passive rote question-and-answer format. This underscores a desire for companion robots with a robust repertoire of emotional expressions, such as abundant non-verbal behaviors. For instance, P3 articulated: *"There are many modes of support, including a comforting smile, a warm embrace, a welcoming gesture, or a smoothing smile like a psychotherapist's...Sometimes talking doesn't solve the problem, comforting through other ways might be better... (for example) noticing that I am feeling down, quietly handing me a glass of water."* Secondly, we observed that some participants exhibited a concern regarding the level of understanding and alignment between their own values and traits with those of their emotional companions. They believe that a greater compatibility in the expression of values fosters attaining a higher-quality companionship. It suggests that robots need to gradually understand and remember users' personality traits and preferences in values. As P1 mentioned, *"If we don't get along, I don't want to say anything at all."* Furthermore, participants also mentioned a scenario where they sometimes may not be able to clearly and fully express their feelings. Being able to understand their emotional state and point out the unspoken emotions they were feeling would evoke a stronger emotional resonance for them, as explained by P1, *"(The ideal capability is)...enable to express what I have left unsaid and my feelings.* Consequently, participants commented they would lose their desire to share their thoughts and feelings with companion robots if the interaction is not rich, delicate, and flexible enough. As P13 mentioned, *"Just like smart speakers, it's interesting at first, but after getting familiar with its functions, it feels repetitive and patterned... I don't want to use it anymore. Many people treat it as a toy. For example, human laughter varies from a smile to a big laugh, there are many different types, but robots only have a few. It feels fresh the first time you see it... But after seeing it a few times, it doesn't seem interesting anymore. I feel the one you showed me (Kebbi) is also just a toy. Keep it as my company? Or talking to it about psychological topics?...(to be honest) I have no desire to have any in-depth conversations with it.".* Here, the participant challenged the robot that served as the stimuli for this focus group discussion, pointing out that despite being equipped with GPT3.5, its interaction mode remains simplistic and inflexible. The improvement in generative language dialogue capabilities cannot overshadow the shortcomings in other abilities and design, such as context awareness, which result in a relatively crude and insufficiently nuanced interaction experience. Participants also mentioned the positive role of soft surfaces in facilitating physical contact, which contributes to relaxation, healing, and the transmission of companionship. However, apart from very few companion robots such as Paro, the majority of them have a hard surface.

*Differentiated value.* During the focus group discussion, participants frequently emphasized their decision to purchase a companion robot would largely depend on whether it provides novel value that surpasses other electronic or software products, and they highlighted design innovation interaction experience based on hardware differentiation might be an approach, as software differentiation is difficult to sustain. Based on the above analysis, participants perceived that mobile floor-standing robots are most likely to demonstrate differentiated value due to their vast application scenarios enabled by their flexible autonomous mobility, followed by pet-type robots, while desktop robots face the most criticism. Participants believe that the flexible autonomous mobility of floor-standing robots can create numerous appealing application scenarios, such as greeting users at the door upon their return home or providing assistance when users encounter a situation in another room. Notably, participants stress that autonomous mobility, when coupled with sufficient situational intelligence and perception capabilities, can generate significant value, without necessitating the robots to directly "execute" for them. For instance, Participant P2 illustrated, *"If the robot cannot lift older adults when they fall, contacting emergency services quickly is also beneficial."* Participants perceived pet-like companion robots could offer a distinct value proposition compared to other ICT solutions, primarily due to their ability to provide a tactile sensation similar to pet fur and mimic pet-like interactive patterns. To be noted, several participants who expressed willingness to adopt such robots possessed or had previously owned a long-term pet with whom they had a close relationship. However, for those who do not have prior pet ownership experience or interest in pet ownership, it is uncertain whether they perceive the same distinct values as those with prior pet ownership. By contrast with the former two types of robots, nearly every group had participants proactively raised the question to our stimuli, the desktop robot Kebbi, *"What are the differences with a smart speaker?"* They questioned the current design of the robot's body components and mobility, arguing that these features were not fully capitalized to create novel and valuable experiences, thus rendering it indistinct from a smart speaker. Except for smart speakers, participants also compared the utilitarian value of our stimuli robot Kebbi against contemporary social media platforms like Douyin (TikTok) and WeChat, as these platforms also facilitate remote social interaction, voice communication, news updates, health monitoring, and entertainment needs.

*The seamless collaboration with aging-in-community infrastructure and services.* Participants proactively introduced the community-based infrastructure and services, emphasizing the importance of integrating robotic capabilities with these existing frameworks and offerings. Participants shared their knowledge and involvement in the new infrastructure and services that emerged under the Chinese government's national policy "aging-in-place and aging-in-community", such as the community-based older adults' canteen, food delivery for older adults, daycare in the community service center, etc. To maximize the value of companion robots, the participants anticipated that companion robots should integrated seamlessly with these frameworks, rather than functioning as isolated information islands. For instance, P7 mentioned that companion robots could assist in fetching food from community canteens and they should aid in contacting community doctors in case of emergency health issues.

## 5 Discussion

In this study, we conducted six participatory focus groups with 18 older adult retirees to explore their perceptions, attitudes, and role expectations for companion robots. In the discussion section, we summarized our key contributions as takeaways and provided design implications for future research and the design of companion robots for cognitively intact, independent-living older adults in light of prior work. Please note that our participants were retirees aged 50 and above. As mentioned earlier, for the sake of conciseness, we will refer to them as "older adults" in the discussion section.

Challenges in Adopting Companion Robots: An Exploratory Study of Robotic Companionship Conducted with Chinese Retirees 17## 5.1 Key contributions

1) We demonstrate that cognitively intact, independent-living older adults harbor emotional support needs, reinforcing their aspiration for aging in place, thereby partially resolving the ongoing debate within the community regarding whether healthy older adults reject companion robots because of a lack of needs.

2) We identify companionship as a primary function of robots in the minds of our participants, illustrating how this companionship can be conveyed through emotional support and specific services.

3) Our research offers insights into the perception and attitudes of underrepresented older adults in East Asian cultures towards companion robots, particularly in China, where rapid urbanization, the one-child policy, and intricate family ties exacerbate the challenges of positive aging. We contribute to the CSCW community by demonstrating the positive significance of companion robots in helping older adults face the transition from the traditional "three generations under one roof" family support model, specifically in the Chinese social context. adding technology-assisted positive aging. According to Linxen et al., 73% of CHI study findings are based on Western participant samples, representing less than 12% of the world's population[54].

4) We uncover several factors influencing the adoption of companion robots, offering insights into the divergence observed in previous studies, which may stem from individual personality differences, the quality of companionship stimuli, the degree of product value differentiation, etc. Our findings also indicate that participants' skepticism towards companion robots during prior work might be partially attributed to the specific design of the stimuli used, rather than their inherent robotic nature. This finding might explain the ineffective adoption of simplistic robots among older adults, which were originally designed for other populations.

5) Through comparing robotic and human companionship, we undercover five advantageous scenarios for robotic companionship and clarify that the perceived roles of robots are abstracted features from specific relationship dynamics, not direct equivalents of human roles. Importantly, participants express the desire for multiple, flexible roles to coexist and shift within their interactions with robotic companions, highlighting the nuanced expectations and preferences of this population.

## 5.2 Re-positioning companion robots for healthy older adults

*5.2.1 Positive aging of healthy older adults.* Prior studies on companion robots largely focused on older adults with dementia, depression, etc. However, we emphasize the crucial significance of providing emotional support to generally cognitively intact, independent-living older adults, especially within China's unique social backdrop, where assisting the only-child generation in navigating the challenges of new eldercare models and fostering successful active aging is essential.

Firstly, with an aging population, maintaining older adults' health, well-being and independence is a key public health priority around the globe[85]. Moreover, amidst profound socio-economic shifts, China's first generation of only-children, now aged 42-53, along with their parents, are entering or nearing old age. The fragmentation of family structures and rapid urbanization have created a unique aging paradigm for this generation, contrasting sharply with the traditional Chinese family-based eldercare models of previous generations. As a result, self-care has become the first option for all middle-aged and old people[92], and Chinese parents seek to avoid becoming a burden for their adult children.[55]. Furthermore, in a collectivistic culture, such as East Asia, self-expression may neither be particularly encouraged nor viewed positively, which further inhibits older adults' ability to obtain emotional support through the aforementioned ways[46]. While the Chinese government has strengthened formal social support through increased

Manuscript submitted to ACM

social care resources, research by Cheng et al. reveals that informal social support, traditionally provided by children, neighbors, and friends, is significantly lower among the young-old (aged 60 to 74) compared to the older-old (aged 75 and above) residing in the same communities[16]. Given that the one-child policy pilot began in 1971 and culminated in the "universal two-child policy" in 2015, this demographic group will comprise a significant proportion of China's aging population. Consequently, how technology can empower this significant population of older adults to successfully navigate the challenges of transitioning eldercare models and attain positive aging is a topic deserving attention. Last but not least, according to elderly care trajectory model proposed by Woll et al., ideally, all assistive technologies should be introduced to the younger older adults in the Pre-trajectory/ everyday life phase, when their functional ability is still intact so that they know how to operate the technology before they really need it[87].

The lifestyle described by participants in our study markedly differed from the stereotypical lonely and empty old age life, as well as from the specific demographic target group for companion robots, such as older adults with dementia or depression[62], which was in agreement with the observation of some previous studies [49, 63], and also reminded the HCI community again that older adults were not a homogeneous group [80]. It is important to note that although the use of companion robots could indeed reduce loneliness and provide emotional support, alleviation of loneliness may not serve as the primary motivating factor for them to purchase. Given the participants' rich social life, it is not surprising that their feedback on companion robots differs from that of elderly individuals with dementia. According to the research by Hung et al., older adults with dementia find the PARO robot to be "like a buddy" helping them uphold a sense of self in the world[38]. However, our elderly participants clearly believe that they do not lack a sense of self in the world, they are the architects of their own positive aging.

Though all participants expressed openness and positivity toward robot-assisted aging, they were divided into different cohorts in terms of the need for robotic companionship varied. Those who were more eager to obtain robotic companionship believed companion robots could play a positive role in many scenarios, providing more timely emotional responses and interactions, and making them feel more relaxed and joyful. This result was fully aligned with previous research by Lazar et al [49], which revealed older adults did not view robotic pets as an all-encompassing solution to remedy isolation but rather expected the response to their emotions, warm and welcoming reactions, social entertainment, etc. For cohorts that were less eager to obtain robotic companionship, our participants still exhibit a heightened sense of urgency in the pursuit of assistive technology for aging-in-place, and expressed unsatisfactory with the quality of robotic companionship and the lack of distinct value differentiation from other existing commercial products within similar categories, such as smart speakers. This finding is in line with the prior research of Fink et al., which indicated robots will be abandoned if users find they cannot bring new added values [27, 60]. It is noteworthy that older adults have conveyed the perspective that, while software updates provide adaptability, hardware differentiation plays a crucial role in providing unique functionalities and user experiences, such as autonomous movement and humanoid form. Future research could take a step further to investigate the significance of embodiment in offering companionship and distinctive added values for companion robots.

Looking back at the developmental history of companion robots, initially, they were used as substitutes for real animals in animal-assisted therapy [52, 68] for lonely individuals and older adults with dementia, etc. However, to serve cognitively intact, independent-living older adults with a larger population, companion robots are more regarded as consumable products rather than therapy. Our findings might explain why current companion robots were regarded as mismatched to their lives and needs by healthy older adults as mentioned in prior work [50, 51, 57]. Future work of companion robots designed for healthy older adults should consider the re-positioning of companion robots. For healthy older adults who are more eager to adopt robotic companionship, a new value positioning that aligns with the needs of relatively healthy older adults is required, namely, shifting from "alleviating loneliness" to "enhancing daily mood". Moreover, to encourage the adoption of robotic companionship among healthy older adults who may be less inclined, it is essential to establish a distinct value proposition.

*5.2.2 Could be a primary function of robots.* Regarding whether companionship can be the primary function, our participants indicated companionship was considered to be on par with other utility-oriented service functions, such as cooking and cleaning. Hence, it is considered acceptable for a robot to be exclusively dedicated to providing companionship as its primary function or to integrate companionship with other functions.

The conclusion of our study differs from prior work which argued companionship was not attractive enough to serve as the primary function of robots [96]. There are two possible reasons for this divergence. Firstly, the difference in participants' demographics. Participants in our study also showed diverse attitudes toward the adoption of robotic companionship, therefore demographic differences might lead to inconsistency in results. Secondly, the quality of companionship may also have an impact on the attitudes of older adults. It remains uncertain about the quality of companionship displayed by the utilized prototype from the participants' perspective.

*5.2.3 Roles of companion robots.* Many previous studies explored the role positioning of companion robots from users' perspective, our findings are in alignment with prior work regarding the most commonly mentioned roles, namely, assistants or servants, friends, family members, and multiple co-existing roles[53]. Furthermore, we identified these role selections cannot be fully understood based on the positioning of that role in human society, which might explain some of the divergences in previous research regarding whether some intimate roles were expected or not[29, 79], as participants might extract different attributes for the same relationship. Hence, in the context of interpreting human-robot relationships through role positioning, these roles function as probes representing attributes that users seek to emulate from human relationships, particularly those deemed most significant, such as the loyalty attribute reflected by the expected role of family members.

Previous research has shown some divergence regarding whether older adults tend to perceive companion robots as tools or as entities with which they can form meaningful relationships[22, 29, 53]. Our study leans towards the belief that both of these attributes coexist. Older adults acknowledge the aspect of companion robots serving as tools while also remaining open to the possibility of forming relationships with them. However, these relationships are not equivalent to those in human society but rather resemble the relationships with "intimate objects" as mentioned by Lazar et al. [2, 49]. Furthermore, in what appears to be a two-way interaction, only one direction is "authentic", while the other comes from the artificial algorithm, older adults believed that even one-way expression also has a positive effect on emotional comfort. The robot's perceived responsiveness functions like an "upgraded version of a confessional", making the process of emotional expression more effective and enjoyable. Future research could also explore the potential of companion robots to act as therapists.

Prior studies also raised ethical concerns over companionship from robots. Some studies have argued that it is a form of deception [48, 82, 84], while others have suggested that older adults value its utility and do not care whether it is a form of deception [8, 91]. Our research results with healthy older adults are in line with the latter. Healthy older adults tend to treat robotic companionship as a self-directed decision rather than a deception, which is totally align with the work of Lazar et al.[49]. Lazar et al., conducted focus groups with independent living older adults and revealed Older adults believed that they were the active party who dominated the choice of whether to establish a relationship with robots through fiction, rather than the passive party who was deceived. Nevertheless, as Turkle et al. mentioned in their work, individuals initially held a perspective of "better than nothing" when interacting with robots, but as they became more aware of the comparative advantages of robots as companions, their perception might turned to favor robots over human sometimes[81]. Therefore, it is unclear whether prolonged usage will lead to shifts in perspectives, especially considering the development of the illusion and the potential implications of robotic companionship on diminishing real-world social interaction. Subsequent research endeavors could explore the potential changes in attitudes through longitudinal studies.

### 5.3 Design implication on high-quality robotic companionship.

In the findings section, we highlighted the difficulty in achieving high-quality companionship and noted dissatisfaction among participants with existing robotic companions. Drawing from these insights, we suggest the following design implications for researchers and designers of companion robots, aiming to explore and enhance the quality of robotic companionship.

*5.3.1 Design companion robot with context-aware multi-role adaptation.* Based on the findings of perceived roles in section 4.2, we revealed participants expected multiple roles to co-exist, indicating the emphasis of expectation for desired roles might vary along scenarios. According to Olsson et al., awareness is a core concept in CSCW, referring to knowledge of proximity, activities, and characteristics of nearby people, considering potential social opportunities[64]. Companion robots should understand the context and adapt roles dynamically and naturally. As we mentioned in section 5.2.3, most existing studies focus on the overall role of companion robots[22, 53, 79]. Our research indicates the necessity for future researchers and designers to gain a deeper understanding of the application scenarios of companion robots in the real lives of older adults, and to comprehend the characteristics of scenarios and role preferences according to different application scenarios, thereby realizing proactive dynamic adaption.

*5.3.2 Design companion robot personas with LLM.* In section 4.3.3, we underscored that robots need to gradually understand and remember users' personality traits and preferences in values, emphasizing the critical role personalized companion robot personas play in enhancing the quality of companionship experiences. Recently, many users on social media have shared DIY examples, focusing on adjusting personas in large language models for more personalized companionship. This trend has attracted attention in communication studies, inspiring research about combining role-playing with large language models[56]. However, the prompts used for defining complex personas can be extremely complicated, making it a very demanding task for most older adults with average digital literacy. We have noticed that some HCI papers have mentioned the integration of role play with large language models[32], but there has been no research yet on crafting personas for LLMs in the context of companion robots for older adults. Consequently, future research could focus on how seniors can actively engage in the pre-definition process of companion robot personas. Furthermore, given the potential evolution of human-robot relationships over prolonged interactions, investigating whether and how users' persona needs may shift, and the strategies for conducting dynamic iterations, will provide invaluable insights into fostering engaging and enduring human-robot companionship.

*5.3.3 Design companion robot with context-aware multimodal verbal and nonverbal behaviors.* In terms of verbal expression, our findings align with previous work that the rigid or incoherent response inhibits the sense of attachment to robots [23, 93]. Nevertheless, the development of large language models has significantly enhanced robots' ability to understand users' intentions and generate natural and coherent responses in dialogues. Participants also emphasized the importance of non-verbal gestures [94] such as eye gaze [11, 58], body movement [71], and facial expression [31], which is consistent with many previous studies. Previous studies on nonverbal behavior predominantly focused on a single modality. Multimodal behavior of companion robots is still unexplored [72]. Multimodal communication is used not only to provide redundancy to transmit intention but also to change the meaning of one mode of communication by augmenting it with a second mode sometimes, the complex cooperation mechanism of multimodal nonverbal cues making it challenging to reuse the research findings of single modalities or other context directly[34]. Therefore, the design of multimodal behaviors should consider adaptability to different contexts. For example, in intimate interactions, robots may employ more eye gaze and facial expressions; whereas in task-oriented interactions, they may focus more on body movements and verbal guidance. By incorporating contextual adaptability into the design of multimodal behaviors, robots can better align with users' expectations and enhance the overall interactive experience.

*5.3.4 Explore collaboration strategy for older adults' companionship scenario.* Apart from the challenge lies in verbal and nonverbal behavior respectively, our findings also show the importance of the collaboration strategy of robots' capabilities, which is highly relevant to the usage scenarios. Collaboration strategy refers to the interaction framework that could coordinate robots' various capabilities to serve the primary purpose, such as providing companionship. Previous research has investigated collaboration strategies across diverse contexts. For instance, Yamazaki et al. [90] compared human–human and experimental human-robot interaction in museums and exhibitions scenarios, and proposed precise coordination strategies of verbal and non-verbal actions under such scenarios. Despite this, there remains a notable gap in the literature concerning collaborative strategies specifically tailored for providing companionship among healthy older adults, highlighting a potential direction for future research. Olsson et al. summarized nine social design objective categories that can significantly enhance the quality of social interaction. These categories can be considered while exploring potential frameworks [64]. Additionally, future research could also delve into user journey mapping or other systematic navigation methods, investigating the roles that robots' various capabilities can play and their collaborative mechanisms within the companionship scenario. This could aid in achieving more efficient and high-quality companionship experiences, while fully leveraging robots' capabilities.

*5.3.5 Build metrics to measure robotic companionship quality.* Our participants underscored that the perceived quality of companionship could potentially affect their willingness to adopt companion robots, along with their expectations of the robots' roles. However, there is a lack of standardized metrics for evaluating the quality of robotic companionship. While previous research has predominantly utilized loneliness scores to gauge the efficacy of companion robots [4, 28, 68], such metrics are limited in scope. They primarily serve as indicators of effectiveness rather than offering a comprehensive framework for comparative analysis of different companion robots on a standardized and modular scale. Consequently, the absence of an evaluative standard has resulted in the inability to quantitatively assess the companionship quality provided by the robots in existing literature. This deficiency complicates efforts to compare and contrast outcomes across studies systematically. Rose et al.[70] proposed a conceptual and methodology framework for robot believability that could serve the empirical studies in this field, future work in companion robots could also explore the potential framework of robotic companionship and possible metrics to evaluate the quality of companionship at a modular level.

## 6 Limitations and future work

This paper presents an exploratory preliminary study, hence, the recruitment did not take into account the balance of many demographic characteristics, such as socioeconomic status, digital literacy, etc. Most participants were retirees from government and large-scale enterprises with middle to high socioeconomic status. They lived in a Tier 1 city (big city) in China, which might have the potential to become early adopters of companion robots. Older adults living in rural areas or with lower economic conditions might hold different attitudes toward robotic companionship. Therefore, more research is warranted to validate the findings of this study with a more diverse population.

We presented different types of companion robots with photos printed on a piece of paper when seeking their preferences for form factor and used one physical desktop humanoid robot as an example in limited study time. Their opinions might be different if they had experiences using companion robots or robots in general.

## 7 Conclusion

To understand healthy older adults' perceptions and attitudes toward companion robots, we conducted multiple participatory focus groups with 18 older adult retirees. Our research provides a view of healthy older adults' perceptions and attitudes toward robotic companionship, the factors that influence the adoption of companion robots, and their views regarding their relationship with companion robots with the probe of role positioning and discussion on ethical concerns. Our research reveals the mismatch between current companion robots' primary value of reducing loneliness with the healthy older adults' need to improve daily mood through immediate emotional interaction. We argue the repositioning of companion robots' value is necessary, if the aim changes from a therapy to a commercial product serving healthy older adults. Besides, older adults exhibit an open and positive attitude toward robots, and the perceived adopting barriers are more related to the unsatisfactory quality of companionship and the unclear differentiation of added value, rather than ethical concerns. Future works include designing metrics and criteria for evaluating the quality of robotic companionship.